\documentclass[%
 reprint,
superscriptaddress,
showpacs,preprintnumbers,
 amsmath,amssymb,
 aps,
]{revtex4-1}

\usepackage{graphicx}
\usepackage{dcolumn}
\usepackage{bm}

\newcommand\redout{\bgroup\markoverwith{\textcolor{red}{\rule[.5ex]{2pt}{0.4pt}}}\ULon}

\usepackage[colorlinks=true,citecolor=blue]{hyperref}
\hypersetup{colorlinks=true,citecolor=blue,linkcolor=red,urlcolor=blue}

\usepackage[colorlinks=true,citecolor=blue]{hyperref}

\def\be{\begin{equation}}
\def\ee{\end{equation}}
\def\bs{\begin{split}}
\def\es{\end{split}}
\def\ber{\begin{eqnarray}}
\def\eer{\end{eqnarray}}

\def\kv{{\bf k}}
\def\qv{{\bf q}}

\def\rv{{\bf r}}

\begin{document}
\renewcommand{\thefigure}{\arabic{figure}}

\title{Relaxation times and charge conductivity of silicene}
\author{Azadeh Mazloom}
\affiliation{Department of Physics, Institute for Advanced Studies
in Basic Sciences (IASBS), Zanjan 45137-66731, Iran }
\author{Fariborz Parhizgar}
\affiliation{School of Physics, Institute for Research in
Fundamental Sciences (IPM), Tehran 19395-5531, Iran}
\author{Saeed H. Abedinpour}
\email{abedinpour@iasbs.ac.ir}
\affiliation{Department of Physics, Institute for Advanced Studies
in Basic Sciences (IASBS), Zanjan 45137-66731, Iran }
\author{Reza Asgari}
\affiliation{School of Physics, Institute
for Research in Fundamental Sciences (IPM), Tehran 19395-5531,
Iran}
\affiliation{Condensed Matter National Laboratory, Institute
for Research in Fundamental Sciences (IPM), Tehran 19395-5531,
Iran}
\date{\today}

\begin{abstract}
We investigate the transport and single particle relaxation times of silicene in the presence of neutral and charged impurities. The static charge conductivity is studied using the semiclassical Boltzmann formalism when the spin-orbit interaction is taken into account. The screening is modeled within Thomas-Fermi and random phase approximations. We show that the transport relaxation time is always longer than the single particle one. Easy electrical controllability of both carrier density and band gap in this buckled two-dimensional structure makes it a suitable candidate for several electronic and optoelectronic applications. In particular, we observe that the dc charge conductivity could be easily controlled through an external electric field, a very promising feature for applications as electrical switches and transistors. Our findings would be qualitatively valid for other buckled honeycomb lattices of the same family, such as germanine and stanine.
\end{abstract}
\maketitle

\section{Introduction}\label{sect:intro}
\begin{figure}
\includegraphics[width=0.9\linewidth]{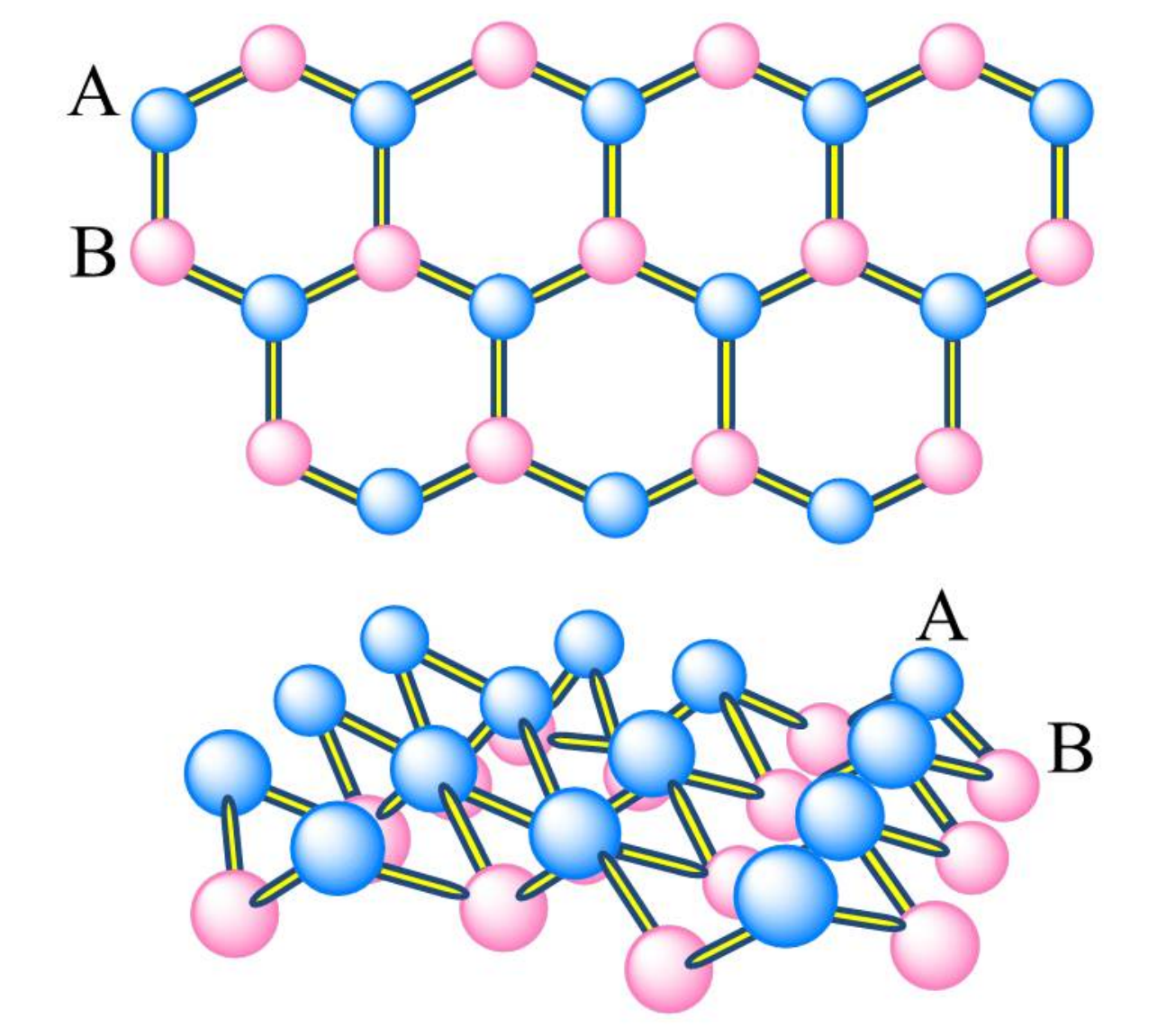}
\caption{ (Color online) Top view (top) and side view (bottom) of the buckled honeycomb structure of silicene, with two sublattices \textit{A} and \textit{B}.}\label{fig:lattice}
\end{figure}

Silicene, a single layer of silicon atoms arranged in a honeycomb structure was foreseen theoretically~\cite{Takeda}. Although silicene was predicted to be stable in its freestanding fashion, so far its existence has been evidenced only on supporting templates on metallic substrates both as a sheet~\cite{Lalmi,Vogt} and as ribbons~\cite{Padova,Aufray} and also on semiconductor substrates~\cite{Fleurence}. Silicon atoms in silicene tend to adopt the tetrahedral $sp^3$ hybridization over the $sp^2$ hybridization of carbon atoms in graphene. This results in a slightly buckled structure of silicene~\cite{Takeda, Durgun, Cahangirov} where atoms in sublattices $A$ and $B$ are displaced from each other in the out-of-plane direction with a distance $d=0.46${\AA} (see, Fig.~\ref{fig:lattice}).
The buckled nature of silicene that originates from the $sp^2/sp^3$ mixed hybridization~\cite{Chen}, together with its large atomic volume, leads to a relatively large spin-orbit coupling (SOC) which is roughly 1000 times larger than graphene~\cite{Min, Yao}, and it is expected to open a visible band gap to be measured at room temperature.
This large SOC is the origin of the envisioned quantum spin-Hall insulator (QSHI) phase, a state with dissipationless spin currents along the edges of the sample, in silicene~\cite{Liu}.
Moreover, a Rashba spin-orbit coupling term also naturally appears in the effective models for silicene, however, this term has a negligible effect on its energy dispersion and could be neglected for most purposes~\cite{Yokoyama, EzawaHall}.

The magnitude of the energy gap in silicene is controllable in different ways, such as photo-irradiation~\cite{Ezawa-ph}, anti-ferromagnetic exchange interactions~\cite{Ezawa-AFM}, and by surface adsorption~\cite{Quhe}.
Also, as the structure of silicene is buckled, application of a perpendicular electric field breaks its sublattice symmetry, and, as a result, a controllable band gap is induced in its energy dispersion~\cite{Ezawa, Drummond}. This perpendicular electric field makes it possible to control the mass term in the Dirac equation independently in valleys $K$  and $K'$. The possibility of controlling the gap term in different valleys, in addition to the strong SOC in silicene, makes valleytronics and spin-valleytronics~\cite{Tahir,EzawaHall,Ezawa-AFM} applications feasible in this material.
The external perpendicular electric field could also spin-polarize the states of each valley. This could be useful in silicene based spin-filtering~\cite{Tsai}.
Remarkably, a transition from a two-dimensional topological insulator state to a band insulator is also predicated through increasing the strength of the electric field~\cite{Ezawa}.
The large spin-orbit coupling in silicene in comparison with graphene, together with the tunability of its band dispersion in particular, makes it a very good candidate to be used in the field of spintronics.

In this paper, we use the semiclassical Boltzmann formalism, within the relaxation time approximation (RTA), to study the charge conductivity of a silicene sheet, subjected to a perpendicular electric field. The effects of both short-range and charged scatterers on the conductivity are investigated.
The Rashba spin-orbit coupling in intrinsic silicene and germane is very tiny and therefore, it is usually neglected in literature. However, one can think of enhancing this term by adding heavy adatoms~\cite{Weeks,Neto,Ma}, or by placing the layer on special substrates such as Ni(111)~\cite{Dedkov,Varykhalov,Jin}. For this purpose, we keep the Rashba SOC throughout our calculations.
Charged impurities are naturally screened by the medium and we take care of this within two different schemes, namely, the Thomas-Fermi approximation (TF) and random phase approximation (RPA). We calculate the static density-density response function~\cite{Giuliani_and_Vignale} in the presence of the SOC term numerically and illustrate how Rashba SOC affects it.
We also explore two different momentum relaxation times, namely, the transport relaxation time and the single particle relaxation time. The former determines the charge conductivity and other transport properties, while the latter shows the quantum-level broadening~\cite{Dassarma}. The effects of both short-range and screened charged impurities are investigated.
Finally, we use the semiclassical Boltzmann formalism to find the charge conductivity of silicene from the transport relaxation time.

The organization of our paper is as follows.
In Sec. \ref{sec:theory}, we briefly describe our model Hamiltonian, together with its band dispersion and eigenfunctions in the presence of Rashba spin-orbit coupling. Then, we describe the density-density response function both at zero and finite Rashba spin-orbit coupling limit. This will be the basis to derive the many-body screening of charged impurities. Also in this section, we describe the single particle as well as transport relaxation times, and charge conductivity within the Boltzmann formalism.
Section \ref{sec:result} is devoted to our analytic and numerical results for relaxation times and conductivity of silicene, in the presence of charged long-range as well as neutral short-range scatterers. We summarize and conclude our main findings in Sec. \ref{sec:summary}.

\section{Model Hamiltonian and Formalism}\label{sec:theory}
The effective low-energy Hamiltonian of the buckled honeycomb lattice of two-dimensional silicene, in the presence of a perpendicular electric field, can be written as~\cite{Ezawa-NJP, EzawaHall, Ezawa-PRB}
\be\label{eq:hamil}
\begin{split}
{\cal H}^\eta=& \hbar v_{\rm F}( k_x \hat{\tau}_x -  \eta k_y\hat{\tau}_y )- \eta \Delta_{\rm SO} \hat{\tau}_z \hat{\sigma}_z   \\
&+ \eta  a_0\lambda_{\rm R}  \hat{\tau}_z(k_x\hat{\sigma}_y - k_y\hat{\sigma}_x) +\Delta_z\hat{\tau}_z~,
\end{split}
\ee
where $v_{\rm F}$ is the Fermi velocity and $a_0$ is the equilibrium lattice constant of silicene.
$k_x$ and $k_y$ are ,respectively, the $x$ and $y$ components of the momentum,
$\hat{\tau}_i$ and $\hat{\sigma}_i$ (with $i=x,y,z$), are the Pauli matrices in the sublattice and spin spaces, respectively, and $\eta=\pm1$ refers to two different valleys $K$ and $K'$. $\Delta_{\rm SO}$ and $\lambda_{\rm R}$ in Eq.~\eqref{eq:hamil} are the intrinsic and Rashba spin-orbit couplings, respectively.
Owing to the buckled structure of the system, atoms of two sublattices $A$ and $B$ are located in two different planes, separated by $d$. Therefore, application of a perpendicular electric field $E_z$ to the sample plane, breaks the sublattice symmetry and splits their on-site energies by $\Delta_z=\pm E_z d/2$. This leads to the splitting of the band dispersion into four subbands in each valley.

The parameters of the Hamiltonian~\eqref{eq:hamil} are $a_0=3.86$ \AA, $d=0.46$ \AA, $v_{\rm F}=5.5 \times 10^5$ m$/$s, $\Delta_{\rm SO}=3.9$ meV and $\lambda_{\rm R}=0.7$ meV.
Note that the Hamiltonian~\eqref{eq:hamil} could be used to model other buckled honeycomb structures such as germanine and stanene, with larger energy gaps of $\Delta_{\rm SO}=23.9$  and $300$ meV, respectively. Material specific parameters of model Hamiltonian~\eqref{eq:hamil} are listed in Ref.~[\onlinecite{Liu}].

The eigenvalues of Hamiltonian~\eqref{eq:hamil} in valley $\eta$ are
\begin{equation}\label{eq:energy}
\varepsilon^\eta_{\beta,s}(k)=\beta \sqrt{(\hbar v_{\rm F} k)^2+\left(\sqrt{\Delta_{\rm SO}^2+(a_0 \lambda_{\rm R} k)^2}+s\eta \Delta_z\right)^2}~,
\end{equation}
where
$\beta=\pm 1$ refers to the conduction and valence bands, and $s=\pm 1$ distinguishes between the light and heavy spin subbands of the conduction and valence bands.
At zero electric field, two subbands are degenerate and a small direct energy gap equal to $2\Delta_{\rm SO}$ opens up between the valence and conduction bands at $k=0$.
Applying a finite gate voltage, splits the two bands into four subbands with the energy gaps of $2|\Delta_{\rm SO}\pm \Delta_z|$.
At a critical value of the electric field $E^{\rm cr}_z=2 \Delta_{\rm SO}/d$, the smaller gap vanishes, and the light subbands become massless with graphene-like linear dispersions $\pm \hbar v_{\rm F} k$ at small wave-vectors.
Increasing the gate field further, the smaller gap opens up once again and increases with $E_z$.
Note that Hamiltonian~\eqref{eq:hamil} describes a two-dimensional topological insulator for $E_z<E_z^{\rm cr}$, and a band insulator for $E_z>E_z^{\rm cr}$.~\cite{Ezawa-NJP}

In typical intrinsic buckled honeycomb lattices such as silicene, germanine, and stanene, the strength of Rashba spin-orbit coupling $\lambda_{\rm R}$ is much smaller than the intrinsic one, $\Delta_{\rm SO}$, and therefore the effects of the Rashba term on the energy dispersion are negligible, even at $k \sim 1/a_0$, which is much beyond the low-energy (\textrm{i.e.}, small $k$) validity region of Hamiltonian~\eqref{eq:hamil}.
Although one can safely discard the Rashba term for most practical purposes and work with much simpler dispersions of
$\pm \sqrt{(\hbar v_{\rm F} k)^2+\left(\Delta_{\rm SO}+s\eta \Delta_z\right)^2}$, as it is possible to enhance the strength of this term by adding heavy adatoms or using special substrates~\cite{Schliemann}, unless otherwise stated, we will retain the Rashba SOC term across our calculations.

\begin{figure}
\includegraphics[width=1.0\linewidth]{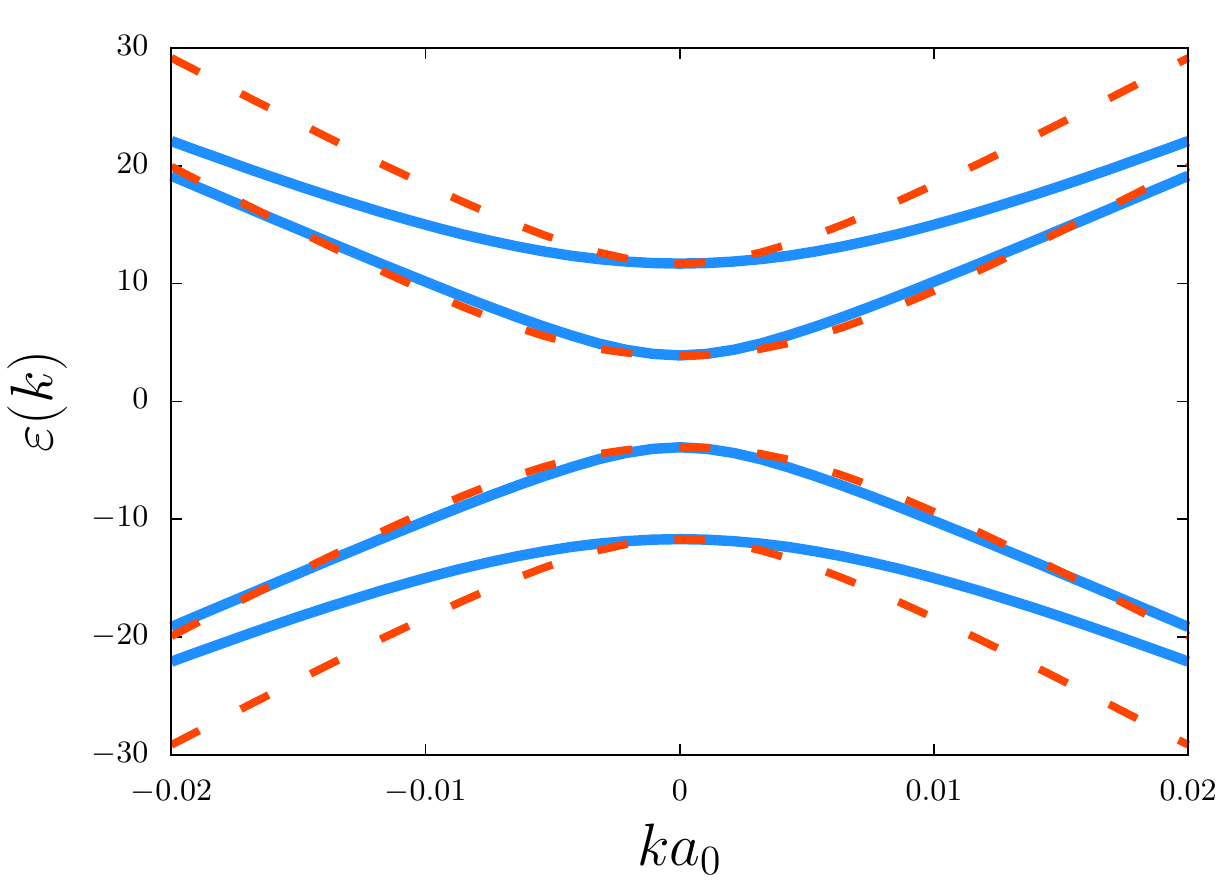}
\caption{ (Color online) Electronic band dispersion of silicene in the K-valley in units of meV, with (dashed lines) and without (solid lines) Rashba spin-orbit coupling. Here the perpendicular electric field $\Delta_z=2\Delta_{\rm SO}$, and the strength of the Rashba term $\lambda_{\rm R}$ has been multiplied by 1000 to magnify its effects on the band dispersion.}\label{fig:disp}
\end{figure}

The low-energy dispersions~\eqref{eq:energy} of silicene are presented in Fig.~\ref{fig:disp}.
Here the effects of Rashba SOC are exaggerated, enhancing  $\lambda_{\rm R}$, 1000 times. Note that band dispersions with the intrinsic value of $\lambda_{\rm R}$ would be essentially indistinguishable from the zero Rashba ones.

The eigenvectors of Hamiltonian~\eqref{eq:hamil}, in the  $\psi=(\psi_{A\uparrow},\psi_{B\uparrow},\psi_{A\downarrow},\psi_{B\downarrow})^T$ basis, could be written in a very compact form as

\be\label{eq:wave}
\psi_{\beta,s}(k,r)=\frac{e^{i\kv\cdot \rv}}{\sqrt{2A}}
\begin{pmatrix}
\beta\, \zeta^\beta(\theta^{s}) \zeta^{s}(\gamma) \\
e^{-i\eta \phi_k} \zeta^{-\beta}(\theta^{s})\zeta^{s}(\gamma)  \\
i \,s\, \beta \, e^{i\eta \phi_k}   \zeta^\beta(\theta^{s})\zeta^{-s}(\gamma)\\
i\, s\, \zeta^{-\beta}(\theta^{s})\zeta^{-s}(\gamma)
\end{pmatrix},
\ee
where $A$ is the sample area, $\phi_k=\arctan(k_y/k_x)$, and $\zeta^+$ and $\zeta^-$ are short-hand notes for the trigonometric functions sine and cosine, respectively. Moreover, we have introduced new angles
\be
\gamma=\frac{1}{2}\arctan{\left(\frac{ a_0 \lambda_{\rm R} k}{\Delta_{\rm SO}}\right)}~,
\ee
and
\be
\theta^\pm= \frac{1}{2} \arctan{\left(\frac{\hbar v_{\rm F} k}{\delta^\pm_k}\right)}~,
\ee
with $\delta^\pm_k= \Delta_z\pm\sqrt{\Delta_{\rm SO}^2+(a_0 \lambda_{\rm R} k)^2}$.

\subsection{Density-density response function and dielectric function}\label{sec:suscep}
In order to  obtain the relaxation times and the electrical conductivity in a system with charged impurities, one needs to know how these impurities are screened in the medium. In this section, we discuss the noninteracting density-density response function of silicene, which will be used to account for the screening of the electron-charged impurity scattering within the RPA.
The noninteracting density-density response function could be obtained from~\cite{Giuliani_and_Vignale}
\be\label{eq:chi}
\chi_0(q,\omega)=\frac{2}{A}\sum_{j,j',\kv}
\frac{n[\varepsilon_{j}(k)]-n[\varepsilon_{j'}(k')]}{ \hbar\omega +\varepsilon_{j}(k)-\varepsilon_{j'}(k') +i 0^+}{\cal F}_{j,j'}(\kv,\kv') ~,
\ee
where $\kv'=\kv+\qv$, $n(\varepsilon)$ is the Fermi-Dirac distribution function, the prefactor $2$ comes from the valley degeneracy, and we use collective indexes: $j=(\beta,s)$ and $j'=(\beta',s')$ for bands and spin subbands.
The form-factor ${\cal F}_{j,j'}(\kv,\kv')$ is the overlap between two wave-functions
\be\label{eq:form}
\begin{split}
 {\cal F}_{j,j'}(\kv,\kv')
&=|<\psi_{j}(\kv)|\psi_{j'}(\kv')>|^2\\
&=\frac{1}{4}\left(1+ \beta \beta' {\cal G}_{j,j'}\right)\left(1+s s' {\cal H}_{j,j'}\right)~,
\end{split}
\ee
where
\be\label{eq:g}
{\cal G}_{\beta,s;\beta',s'}(\kv,\kv')\equiv \frac{\hbar^2 v_{\rm F}^2\kv\cdot\kv'+\delta^s_k \delta^{s'}_{k'}}{|\varepsilon_{\beta, s}(k) \varepsilon_{\beta', s'}(k')|}~,
\ee
and
\be\label{eq:h}
{\cal H}_{\beta,s;\beta',s'}(\kv,\kv') \equiv \frac{\Delta_{\rm SO}^2+a^2_0 \lambda^2_{\rm R}  \kv\cdot \kv'}{|\varepsilon_{\beta, s}(k) \varepsilon_{\beta', s'}(k')|}~.
\ee
Using Eqs. \eqref{eq:form}-\eqref{eq:h} in Eq.~\eqref{eq:chi}, one can numerically calculate the polarizability of silicene with the Rashba spin-orbit coupling, and in the presence of a perpendicular electric field.

\emph{Density-density response function at the zero Rashba limit.}
As we have already pointed out, the intrinsic Rashba spin-orbit coupling is very small, and it is often reasonable to neglect it for simplicity~\cite{Chang, Tabert}.
Taking $\lambda_{\rm R}=0$, the Hamiltonian~\eqref{eq:hamil} becomes diagonal in the spin space,
\be\label{eq:hamil_gapped}
{\cal H}^{\eta,s}(k)= \hbar v_{\rm F}( k_x \hat{\tau}_x -  \eta k_y\hat{\tau}_y )+\left(\Delta_z+s \eta \Delta_{SO} \right)\hat{\tau}_z~,
\ee
which is identical to the Hamiltonian of a gapped graphene with spin dependent gap values $\Delta_\pm=|\Delta_{\rm SO} \pm \Delta_z|$ for each spin component~\cite{Pyatkovskiy}.
In other words, the $z$-component of the spin is conserved, and the external electric field simply splits two spin sub-bands. Note that as the time reversal symmetry is still preserved, the direction of spin splitting should be reversed between two valleys.

The noninteracting density-density response of this simplified model becomes
$\chi_0(q,\omega)=\chi^{\rm g}_0(q,\omega,\Delta_+)+\chi^{\rm g}_0(q,\omega,\Delta_-)$, where $\chi^{\rm g}_0(q,\omega,\Delta)$ is the noninteracting response function of a gapped graphene with gap $\Delta$.
This quantity is analytically known, both along the real~\cite{Pyatkovskiy} and imaginary~\cite{Qaiumzadeh} frequency axes, and in the static (i.e., $\omega=0$) limit reads
\be\label{eq:chi_g_1}
\begin{split}
\chi^{\rm g}_0&(q,\Delta)=-\nu_{\rm g}(\mu)\left[1-\Theta(q-2k_{\rm F})\left(\frac{\sqrt{q^2-4k_{\rm F}^2}}{2q}\right. \right.\\
&\left.\left.-\frac{(\hbar v_{\rm F} q)^2-4\Delta^2}{4 \hbar v_{\rm F} q\mu}\arctan(\frac{ \sqrt{(\hbar v_{\rm F} q)^2-4(\hbar v_{\rm F} k_{\rm F})^2}}{2\mu})\right)\right]~,
\end{split}
\ee
for $\mu>\Delta$, and
\be\label{eq:chi_g_2}
\chi^{\rm g}_0(q,\Delta)=
-\frac{1}{\pi \hbar^2 v_{\rm F}^2}
\left[\Delta
+\frac{(\hbar v_F q)^2 -4\Delta^2}{2\hbar v_{\rm F} q}\arctan\left(\frac{\hbar v_{\rm F} q}{2\Delta}\right)\right]~,
\ee
for $\mu<\Delta$.
In Eqs.~\eqref{eq:chi_g_1} and~\eqref{eq:chi_g_2}, $\mu$ is the chemical potential, $k_{\rm F}=\sqrt{\mu^2-\Delta^2}/(\hbar v_{\rm F})$, and
$\nu_{\rm g}(\varepsilon)=2 \varepsilon \Theta(\varepsilon-\Delta)/(\pi \hbar^2 v_{\rm F}^2)$ is the density of states (per unit area) of the gapped graphene, where $\Theta(x)$ is the Heaviside step function. Note that the spin degeneracy of graphene is absent here.
\begin{figure}
\includegraphics[width=1.0\linewidth]{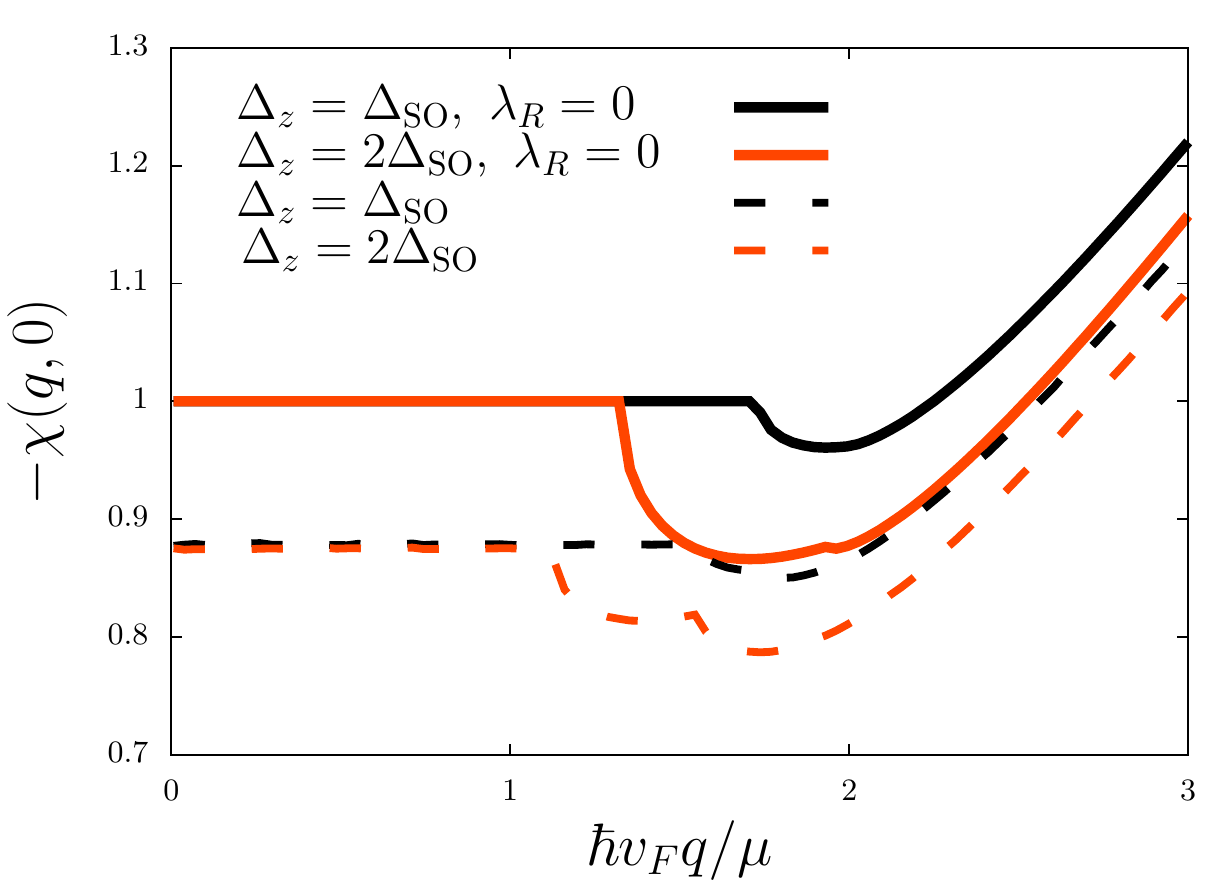}
\caption{ (Color online) Static density-density response function of silicene [in units of $4 \mu/(\pi \hbar^2 v_{\rm F}^2)$] as a function of wave-vector $q$ [in units of $\mu/(\hbar v_{\rm F})$], in the presence (dashed lines) and absence (solid lines) of Rashba spin-orbit coupling, and for two different values of the external electric field $\Delta_z$. Here again we have multiplied the Rashba term $\lambda_{\rm R}$ by 1000 to magnify its effects on the polarizability.
\label{fig:xi}}
\end{figure}

The noninteracting density-density response function of silicene is presented in Fig. \ref{fig:xi}.
The analytic expressions of Eqs.~\eqref{eq:chi_g_1} and~\eqref{eq:chi_g_2} are compared with the full numerical evaluation of Eq.~\eqref{eq:chi} at non-zero Rashba SOC and for two different values of the external electric field.  Similar to Fig.~\ref{fig:disp}, we enhance the value of $\lambda_{\rm R}$, 1000 times to illustrate its effect on the polarizability.
In both cases, $\chi_0(q)$ is constant and equal to $-\nu(\mu)$ at low wave-vectors, where $\nu(\mu)$ is the density of states (per unit area) of silicene at its Fermi level. At larger values of the wave-vector, on the other hand, the absolute value of the response function increases linearly with $q$.

Once the noninteracting density-density response function is known, one can obtain the screened interaction between electrons and charged impurities $V^{\rm sc}_{\rm e-i}(q)=V_{\rm e-i}(q)/\epsilon(q)$, where $\epsilon(q)$ is the dielectric function, and within the RPA reads $\epsilon^{\rm RPA}(q)=1-v_q\chi_0(q)$,
where $v_q=e^2/(2\epsilon \epsilon_0 q)$ is the Fourier transform of the electron-electron interaction $e^2/(4\pi \epsilon \epsilon_0 r)$ in two dimensions ($2D$), with $\epsilon_0$ being the vacuum permittivity, and $\epsilon$ the dielectric constant of the environment, which we take to be equal to 1 throughout this work.
Replacing $\chi_0(q)$ in the RPA dielectric function with its long-wavelength limit results in the well-known Thomas-Fermi approximation for screening $\epsilon^{\rm TF}(q)=1+\nu(\mu)v_q$.

\subsection{Relaxation times and conductivity}
Within the relaxation time approximation, the transport scattering time of an electron  in the $j$-th band and with wave-vector $k$, could be written as~\cite{singh_book}
\be\label{eq:taut}
\frac{1}{\tau^j_{\rm tr}(\kv)}=\frac{1}{A}\sum_{j',\kv'} W_{j,j'}(\kv,\kv') \left[1-\frac{v_{j'}(\kv')}{v_j(\kv)}\cos(\theta_{\kv,\kv'})\right]~,
\ee
where the summation is over all possible final states with band $j'$ and wave-vector $\kv'$, $v_j(k)$ is the group velocity of band $j$, and $W_{j,j'}(\kv,\kv')$ is the transition probability between states $(j,\kv)$ and $(j',\kv')$, which , by using the Fermi's golden rule for elastic scatterings, could be written as
\be
W_{j,j'}(\kv,\kv')=\frac{2\pi}{\hbar}\left|T_{j,j'}(\kv,\kv')\right|^2 \delta(\varepsilon_{j',k'}-\varepsilon_{j,k})~.
\ee
Here $T_{j,j'}(\kv,\kv')$ is the $T$-matrix~\cite{mahan_nutshell}, and within the first Born approximation for dilute and randomly distributed impurities becomes proportional to the matrix elements of the scattering potential~\cite{kohn_luttinger}
\be
\left|T_{j,j'}(\kv,\kv')\right|^2 \approx n_{\rm imp} {\cal F}_{j,j'}(\kv,\kv') V^2_{\rm e-imp}(\kv-\kv')~,
\ee
where $n_{\rm imp}$ is the density of impurities and $V_{\rm e-imp}(\qv)$ is the Fourier transformation of the interaction potential between an electron and a single impurity, and depends on the nature of impurities. In the case of short-ranged non-magnetic impurities (e.g., defects or neutral adatoms) it is quite safe to approximate it with a zero-range hard-core potential $V_{\rm e-imp}(\qv)=V_0$.
On the other hand, scattering from charged impurities will be long-ranged Coulombic, $V_{\rm e-imp}(\qv)\sim 1/q$. This potential is screened by other electrons of the system. The screening could be treated within the {\textit{e.g.}, Thomas-Fermi or random-phase approximations.~\cite{Giuliani_and_Vignale}

Energy conservation in an elastic scattering prevents any transition between valence and conduction bands.
In the zero Rashba limit, $\sigma_z$ commutes with Hamiltonian~\eqref{eq:hamil}, and spin is a good quantum number. Therefore, with spin-independent scatterers such as the ones we consider in this work, the transition between two spin subbands of the same (valance or conduction) band will be prohibited too. The diagonal components of the form-factors in this case will read
\begin{equation}
{\cal F}_{\beta,s;\beta,s}(\kv,\kv')=\frac{1}{2}\left(1+\frac{\hbar^2 v^2_{\rm F}\kv\cdot \kv'+\Delta^2_s}{\varepsilon_{k,s}\varepsilon_{k',s}}\right)~,
\end{equation}
where $\Delta_s=|\Delta_{\rm SO}+s\Delta_z|$ denotes the gap between $s$-spin subbands, and $\varepsilon_{k,s}=\sqrt{(\hbar v_{\rm F} k)^2+\Delta_s^2}$ is the dispersion of massive Dirac fermions with a mass term of $\Delta_s$.
But the situation is different in the presence of a finite Rashba term. As the $z$-component of the spin of an electron is no longer conserved, inter-subband transitions would be allowed even in the absence of magnetic impurities. However, as we will show below, the rate of this spin-flip transition would be much smaller than the intra-subband transitions, for the intrinsic values of $\lambda_{\rm R}$.

 In addition to the transport relaxation times, one can define single particle relaxation times, which are related to the imaginary part of the single particle self-energies, and define the broadening of the single particle levels owing to the impurity scatterings\cite{mahan_nutshell, Dassarma}:
\be\label{eq:tau_s}
\frac{1}{\tau^j_{\rm sp}(\kv)}=\frac{1}{A}\sum_{j',\kv'} W_{j,j'}(\kv,\kv')~.
\ee

Having calculated the transport relaxation times, the charge conductivity of the system could be obtained using the semi-classical Boltzmann formalism~\cite{mahan_nutshell, Dassarmarev}
\begin{equation} \label{eq:sigma}
\sigma =\frac{e^2\hbar ^2 v_{\rm F}^2}{2} \sum_{j} \int \mathrm{d}\varepsilon \, \nu_j(\varepsilon) \tau_{\rm tr}^j(\varepsilon) \left(-\frac{\partial n(\varepsilon)}{\partial \varepsilon}\right)~,
\end{equation}
where $\nu_j(\varepsilon)$ is the density of states of the $j$-th band.
At zero temperature, $n(\varepsilon)$ is a step function, and we arrive at the renowned expression
\begin{equation}\label{eq:sigma2}
\sigma=\frac{e^2v_{\rm F}^2}{2} \sum_s \nu_s(\mu) \tau_{\rm tr}^s(k_{{\rm F},s})~.
\end{equation}
Naturally, only the bands which cross the Fermi energy contribute to the conductivity at low temperatures.
Note that this semiclassical theory is valid only in the regime of dilute impurity concentration \textit{i.e.}, when the density of impurities is much smaller than the density of charge carriers.

\section{Results and Discussions}\label{sec:result}
In this section, we turn to the presentation of our analytical and numerical results for single particle and transport relaxation times as well as the charge conductivity of silicene in the presence of short-range and long-range charged scatterers.

\subsection{Relaxation times}

\begin{figure}
\centering
\includegraphics[width=1.0\linewidth]{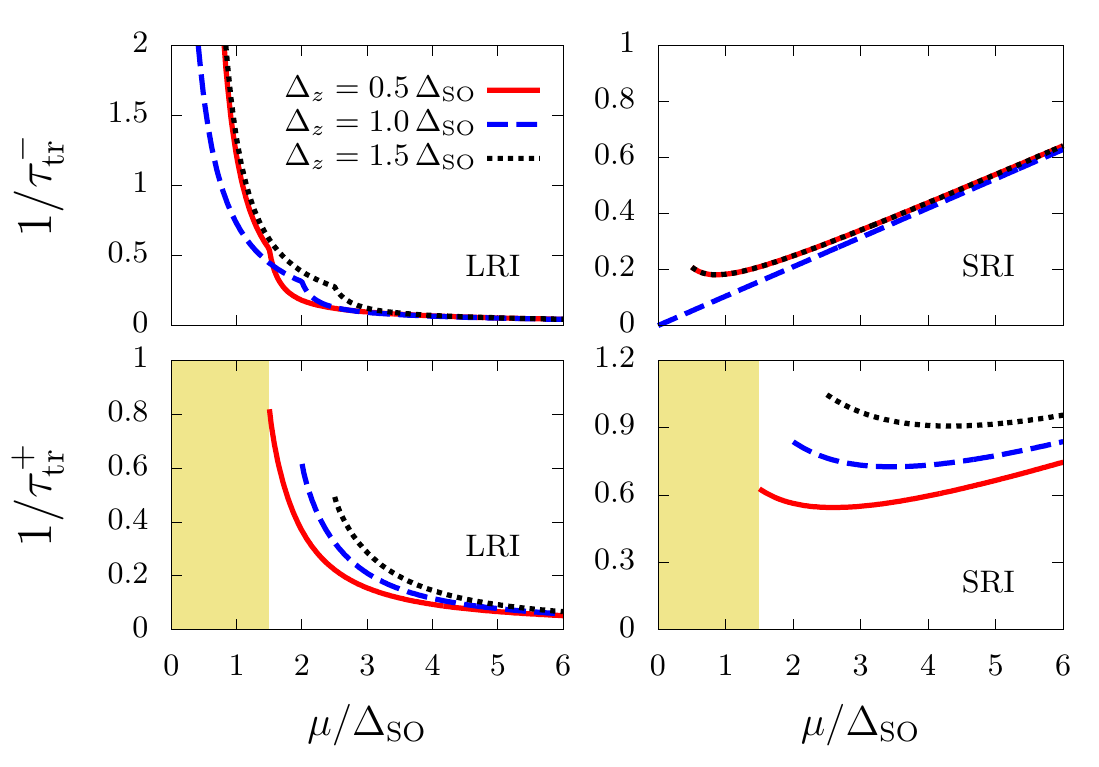}
\caption{(Color online) Inverse of the transport relaxation times of lower (top) and upper (bottom) spin subbands of the conduction band in units of (ps)$^{-1}$ at the Fermi energy, versus chemical potential for different values of the perpendicular electric field $\Delta_z$. $V_0= 3$\,{\rm keV}\AA$^2$ is the strength of short-range potential, and the concentration of both long-range (left) and short-range (right) impurities assumed to be $n_{\rm imp}=10^{9} $cm$^{-2}$, and the long-range potential is screened within RPA. 
Filled areas indicate the energy gap region of the upper subband for $\Delta_z=0.5 \Delta_{\rm SO}$.
\label{fig:taumu}}
\end{figure}

\begin{figure}
\includegraphics[width=1.0\linewidth]{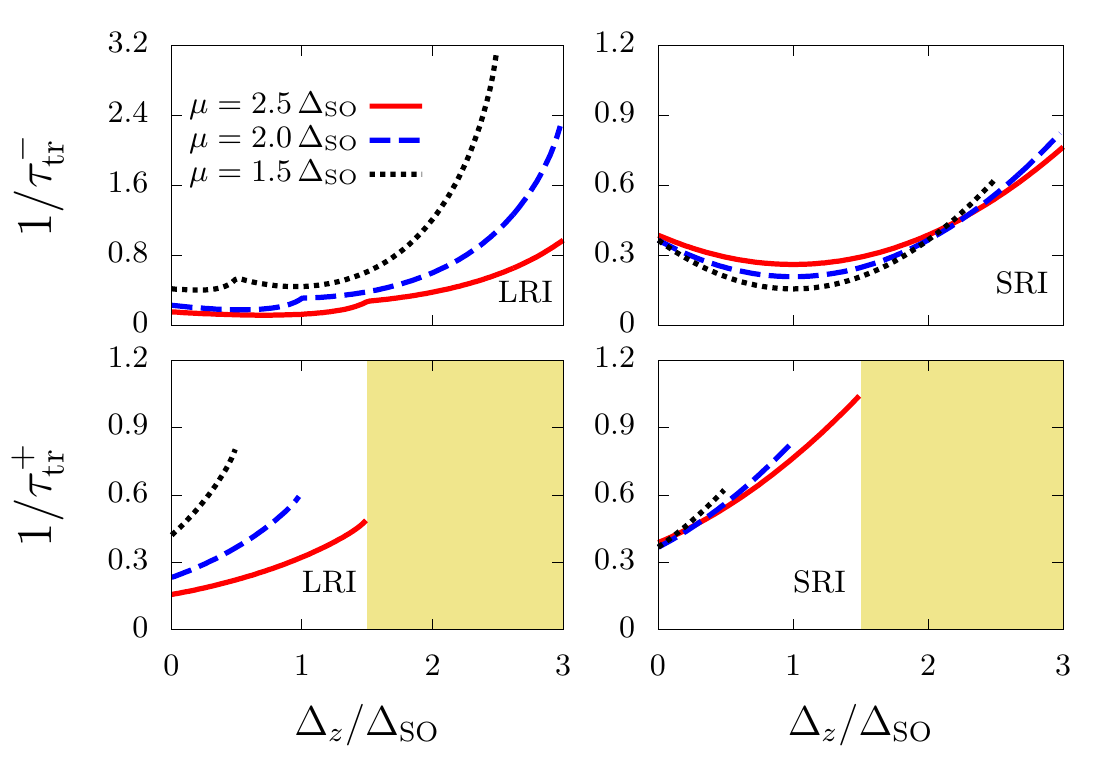}
\caption{(Color online) Inverse of the transport relaxation times of lower (top) and upper (bottom) spin subbands of the conduction band in units of (ps)$^{-1}$ at the Fermi energy, versus perpendicular electric field $\Delta_z$, for different values of the chemical potential. We use $V_0= 3$\, {\rm keV}\AA$^2$ for the strength of short-range potential, and an impurity density of $n_{\rm imp}=10^{9}$ cm$^{-2}$ for both long-range (left) and short-range (right) impurities, and have screened the long-range potential within the RPA.
Filled areas indicate the energy gap region of the upper subband for  $\mu=2.5 \Delta_{\rm SO}$.
\label{fig:tauZ}}
\end{figure}

Relaxation times are generally functions of the incoming electrons' wave-vector, and at low temperatures only states around the Fermi level will contribute to transport and single particle properties. In the following we will concentrate only on the behavior of relaxation times at the Fermi energy.
The transport and single particle relaxation times could be conveniently obtained using Eqs. \eqref{eq:taut} and \eqref{eq:tau_s}, respectively.
In the case of short-range neutral impurities, assuming vanishing Rashba SOC, the corresponding summations over final states could be performed analytically and the final results read
\be
\frac{1}{\tau_{\rm tr}^s(\varepsilon_F)}=\frac{n_{imp} V_0^2}{4\hbar^3 v_F^2\mu}\left(\mu^2+3\Delta_s^2\right)\Theta(\mu-\Delta_s)~,
\ee
and
\be
\frac{1}{\tau_{\rm sp}^s(\varepsilon_F)}=\frac{n_{imp} V_0^2}{2\hbar^3 v_F^2\mu}\left(\mu^2+\Delta_s^2\right) \Theta(\mu-\Delta_s)~.
\ee
For charged impurities, as we already discussed in the previous section, the inclusion of many-body screening is essential. Within the Thomas-Fermi approximation the screened potential can be written as
\begin{equation}\label{eq:v_tf}
V^{\rm TF}_{\rm e-imp}(q)=\frac{V_{\rm e-imp}(q)}{\epsilon_{\rm TF}(q)}=\frac{e^2}{2 \epsilon_0(q+q_{\rm TF})}~,
\end{equation}
where $V_{\rm e-imp}(q)=-e^2/(2 \epsilon_0 q)$ is the bare Coulomb interaction of an electron with an impurity of charge $+e$, and $q_{\rm TF}=\nu(\mu)/(2\epsilon_0) $ is the TF wave-vector.
At the zero Rashba SOC limit, again the relaxation times could be obtained analytically.
The final expressions for single particle and transport relaxation times are quite lengthy and we have presented them in the appendix.
A more accurate screening of the charged impurities would be achieved within the RPA. In this case we present our results for the relaxation times numerically, retaining the intrinsic value of the Rashba spin-orbit coupling in our calculations.
In the following we will use $n_{\rm imp}=10^9 cm^{-2}$ for the impurity concentration of both short range and long range scatterers. This guarantees that the diluteness criteria will be satisfied for a wide range of chemical potentials $\mu$, and perpendicular electric fields $\Delta_z$ in the following results for relaxation times and charge conductivities. Note that in an intrinsic silicene, and in the absence of any perpendicular electric field, even a chemical potential of $\mu \approx 2 \Delta_{\rm SO}$, roughly corresponds to the carrier density of $10^{10} cm^{-2}$, which is already much larger than $n_{\rm imp}$.

Our results for the transport relaxation times of both conduction spin subbands at the Fermi levels, and in the presence of short-range (SR) and long-range (LR) charged impurities, are presented as functions of the chemical potential and the perpendicular electric field in Figs. \ref{fig:taumu} and \ref{fig:tauZ}, respectively. In the case of charged impurities, the electron-impurity interaction is screened within the RPA.
Note that the relaxation time of a band at the Fermi energy is not defined when the Fermi energy does not intersect that band.
In the case of charged impurities a cusp like feature in the relaxation time of the lower ($-$) band as a function of the chemical potential or perpendicular electric field appears when the chemical potential reaches the bottom of the upper ($+$) band. The origin of this behavior lies in the transition between metallic and insulating contributions of the upper bands to the screening of the electron impurity interaction. As the short range interaction is not screened, such a behavior is absent there.

As we already discussed, within the TF approximation for screening, one simply replaces the full wave-vector dependent polarizability $\chi_0(q)$, with its $q\to0$ limit. In 2D systems $\chi_0(q)$ is constant at long wavelengths (see, Fig. \ref{fig:xi}) therefore one would expect a good agreement between RPA and TF.
In Fig. \ref{fig:tauTF_RPA}, we compare our results obtained for the transport relaxation times within these two different screening approximations of the long-range potential.
The only region where there is a visible difference between two different schemes, is the relaxation time of the lower subband at small chemical potentials.
This is precisely the regime where only the lower conduction band is occupied.
The RPA would take into account the contribution of the upper band in screening, using Eq.~\eqref{eq:chi_g_2} for its density-density response function, while the TF approximation will simply ignore it, as this response function vanishes at $q=0$.
\begin{figure}
\includegraphics[width=0.9\linewidth]{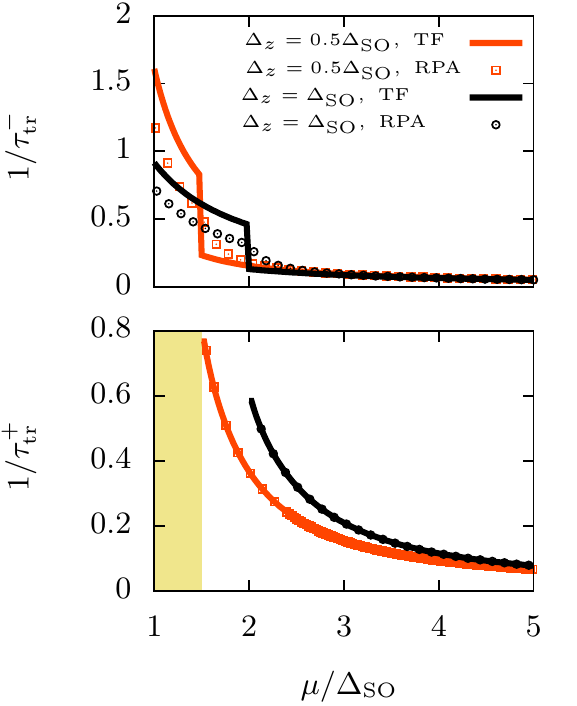}
\caption{ (Color online) Inverse of the transport relaxation time of lower (top) and upper (bottom) conduction bands in units of (ps)$^{-1}$ as a function of the chemical potential $\mu$, in the presence of charged impurities, and for different values of the perpendicular electric field $\Delta_z$. Solid lines refer to analytic results obtained using TF approximation for the screening of the long-range interaction, while symbols refer to numerical results with RPA screening. The concentration of impurity is $n_{\rm imp}=10^{9}$ cm$^{-2}$ and the filled area shows the energy gap region of the upper subband for $\Delta_z=0.5 \Delta_{\rm SO}$.
\label{fig:tauTF_RPA}}
\end{figure}

Now, we turn to the presentation of our results for single particle relaxation times, as introduced in Eq.~\eqref{eq:tau_s}.
Figure \ref{fig:tauts} shows the ratio between transport  and single particle relaxation times, as a function of the chemical potential.
The transport relaxation time is always larger than the single particle one.
For the upper conduction band, the ratio increases with chemical potential, while for the lower one this ratio quickly saturates to $2$. In the case of long-range impurities, abrupt changes in the ratio between transport and single particle relaxation times for lower band again originates from the transition between insulating and metallic screenings of the upper band.

\begin{figure}
\includegraphics[width=1.0\linewidth]{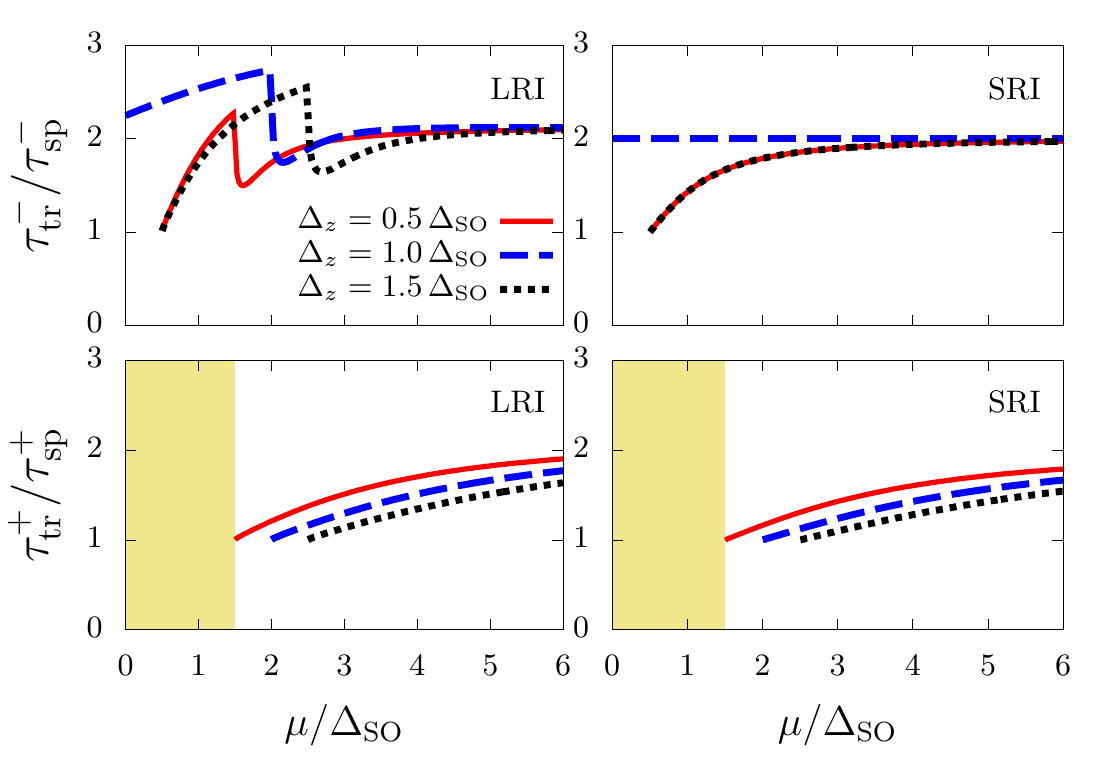}
\caption{(Color online) The ratio between transport and single particle relaxation times at the Fermi level as a function of the chemical potential and for different values of the perpendicular electric field $\Delta_z$.
Top panels are the ratios for the lower conduction bands and bottom panels for the upper conduction bands.
The concentration of both long-range (left) and short-range (right) impurities assumed to be $n_{\rm imp}=10^{9}$ cm$^{-2}$, $V_0= 3$ \,{\rm keV}\AA$^2$ is used for the strength of the short-range potential, and the long range potential is screened within the RPA. The Filled area is the energy gap region of the upper subband for $\Delta_z=0.5 \Delta_{\rm SO}$.
\label{fig:tauts}}
\end{figure}

In the last figure of this section, we assume a finite value for Rashba spin-orbit coupling $\lambda_{\rm R}=0.7$ meV. This off-diagonal term in the spin space permits interband transitions between two conduction spin subbands.
As the intrinsic value of the Rashba spin-orbit coupling is very small in silicene, it still makes sense to label two conduction subbands as up and down spin subbands. In this sense, the interband transition could be interpreted as spin relaxation or spin-flip.
In Fig. \ref{fig:taupm} we present the inverse of the interband or spin-flip single particle relaxation time,
\be\label{eq:tau_sf}
\frac{1}{\tau^{+-}_{\rm sp}(\kv)}=\sum_{\kv'} W_{+,+;+,-}(\kv,\kv')~,
\ee
for the intrinsic value of the Rashba spin-orbit coupling of silicene~\cite{Liu}. This interband relaxation time is much  (almost $10^6$ times) longer than the intraband relaxation times, presented in our previous figures.
\begin{figure}
\includegraphics[width=0.9\linewidth]{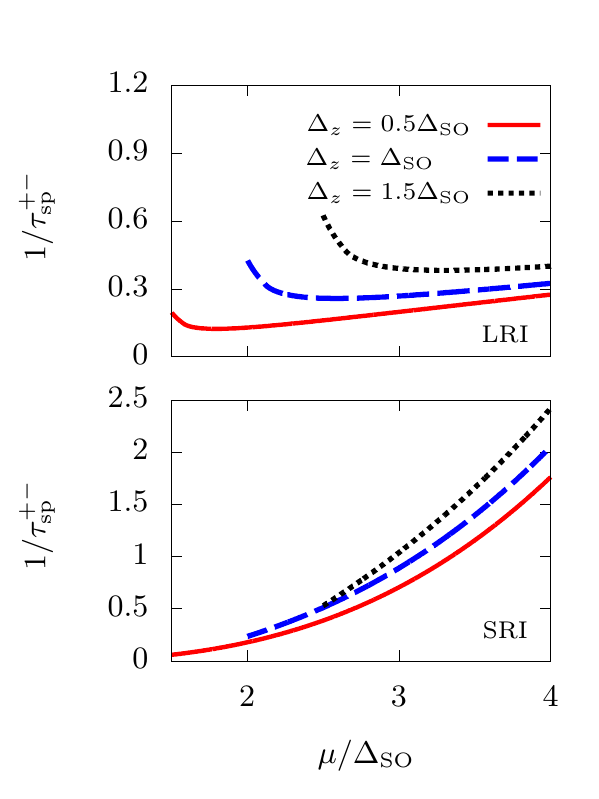}
\caption{ (Color online) Inverse of the interband single particle relaxation time between two conduction spin subbands in units of ($\mu$s)$^{-1}$ as a function of chemical potential and for different values of the perpendicular electric field $\Delta_z$, for charged (top) and neutral (bottom) impurities.
We use an impurity density of $n_{\rm imp}=10^{9}$ cm$^{-2}$ in both cases, $V_0= 3$ \,{\rm keV}\AA$^2$ for the short-range interaction strength, and screen the long-range potential within the RPA.
\label{fig:taupm}}
\end{figure}

\subsection{Charge conductivity}
Now, having calculated the transport relaxation times, we can calculate the charge conductivity using Eq.~\eqref{eq:sigma2}.

Figure \ref{fig:sigmamu} illustrates our results for the charge conductivity of silicene in the presence of long-range and short-range scatterers as a function of the chemical potential for two different values of the perpendicular electric field, \textrm{i.e.}, $\Delta_z=0.5 \Delta_{\rm SO}$, for which both bands are gapped, and for $\Delta_z= \Delta_{\rm SO}$ which makes the lower band gapless.
In the case of long-range impurities, the comparison is made between RPA and TF screenings of the electron-impurity interaction, too. The conductivity is an increasing function of $\mu$, and therefore of the carrier density. In the case of short-range impurities, one can see that at large chemical potentials it saturates to a constant value determined by the scattering strength and impurity concentration [see also Eq.~(\ref{sigma_short})].
As one would expect from the behavior of relaxation times, conductivity jumps when the chemical potential reaches the bottom of the upper band.
This makes silicene suitable for making charge switches. Note that, the conductivity becomes zero when the chemical potential lies inside the gap of both bands.
Interestingly, in the case of short-range scatterers, and for $\Delta_z=\Delta_{\rm SO}$ where the lower band is massless, the conductivity remains constant, as long as $\mu$ has not reached the upper band. This is in agreement with what has been observed in graphene~\cite{Dassarmarev}.

\begin{figure}
\includegraphics[width=1.0\linewidth]{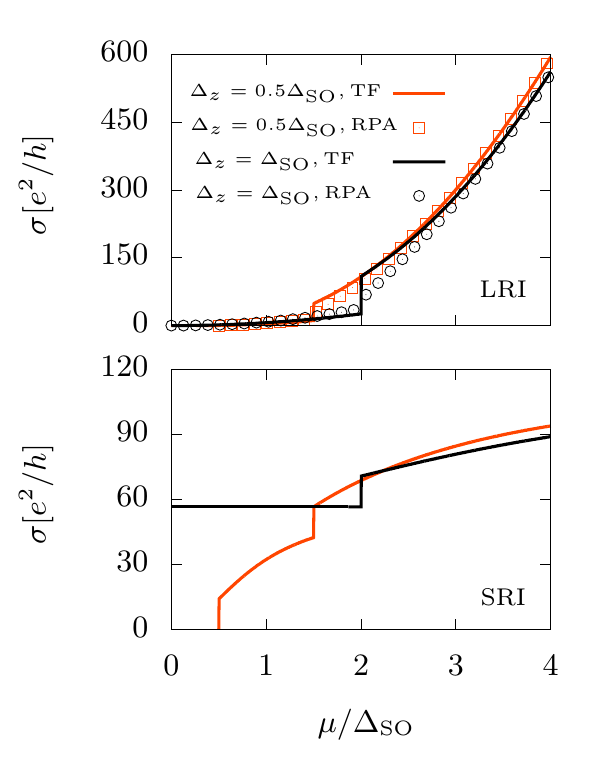}
\caption{(Color online) Boltzmann conductivity of silicene (in units of $e^2/h$) as a function of chemical potential $\mu$ for two different values of the perpendicular electric field, and in the presence of long-range (top) and short-range (bottom) impurities. In the case of long-range impurities, results of both RPA (symbols) and TF (solid lines) screenings are presented.
Similar to previous figures, $V_0=3$\,{\rm keV}\AA$^2$ and $n_{\rm imp}=10^{9}$ cm$^{-2}$ are used.
\label{fig:sigmamu}}
\end{figure}

\begin{figure}
\includegraphics[width=1.0\linewidth]{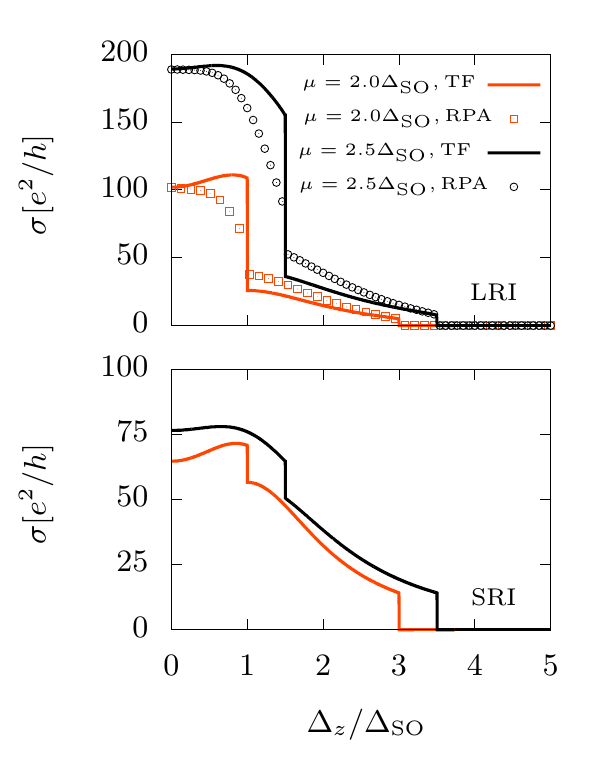}
\caption{(Color online) Boltzmann conductivity of silicene (in units of $e^2/h$) as a function of perpendicular electric field $\Delta_z$ for two different values of the chemical potential, and in the presence of long-range (top) and short-range (bottom) impurities. In the case of long-range impurities, results of both RPA (symbols) and TF (solid lines) screenings are presented. We have also used $V_0=3$\,{\rm keV}\AA$^2$ and $n_{\rm imp}=10^{9} $cm$^{-2}$.
\label{fig:sigmaZ}}
\end{figure}

Finally, in Fig.~\ref{fig:sigmaZ} the dependence of charge conductivity on the perpendicular electric field is illustrated for long-range and short-range scatterings.
A similar behavior to previous figure is also observed here. Increasing the perpendicular electric field, a sudden drop in the conductivity appears in the transition from the double band to the single band regime. Increasing $\Delta_z$ further, conductivity drops to zero, when the lower band is also pushed above the chemical potential.

\section{Summary and conclusions}\label{sec:summary}
In summary, we have investigated the transport properties of a silicene sheet in the presence of both intrinsic and Rashba spin-orbit couplings. Although the intrinsic value of the Rashba spin-orbit coupling is small, it can be enhanced using heavy adatoms or appropriate substrates. Also, we have considered a perpendicular electric field applied to silicene sheet which can be used to modify the band gap and therefore the transport properties of silicene.

Electron-impurity scatterings in two dimensional systems can originate from charged impurities, that mainly come from the substrate or from the local defects. In the latter case, the scattering potential is short range, while in the former it is long-ranged coulombic.
Using an effective low-energy Hamiltonian for silicene in the presence of a perpendicular electric field, we have calculated the transport relaxation time for both types of scatterers. For charged impurities, we have employed Thomas-Fermi and random phase approximations for screening.

We have also compared the single particles relaxation time, which measures the quantum level broadening of the system, with the transport relaxation time. We have found that the transport relaxation time is always larger than the single particle one. Moreover, we have shown that the interband relaxation time is much longer than the interband relaxation time. Having calculated the transport relaxation times, and using the Boltzmann approach for conductivity, we calculated the charge conductivity in the presence of short-range and long-range impurities. As silicene is a material with a tunable band dispersion we have observed that the conductivity can possess abrupt changes with respect to the perpendicular electric field. This makes silicene a good candidate for electronic device applications where the switching process is a need.

\appendix*
\section{Analytical results for relaxation times and conductivities}\label{app}

In the limit of vanishing Rashba SOC, i.e.  $\lambda_{\rm R}=0$, it is possible to obtain analytical expressions for relaxation times and conductivities in the presence of short-range impurities and in the presence of long-range impurities when the TF approximation is used for screening.

\textit{Short-range impurities-}
In the case of short-range impurities, the summation over $\kv'$ in Eq. \eqref{eq:taut} could be performed analytically,
\be
\begin{split}
\frac{1}{\tau^s_{\rm tr}(\varepsilon_F)}&= \frac{n_{imp} V_0^2}{4\pi \mu \hbar^3 v_{\rm F}^2}\Theta(\mu-\Delta_s)\\
&\times \int_0^{2\pi} \mathrm{d}\theta \left(1-\cos\theta\right) \left[\mu^2+\Delta_s^2+\left(\mu^2-\Delta_s^2\right)\cos\theta\right]\\
&=\frac{n_{imp} V_0^2}{4\hbar^3 v_{\rm F}^2\mu}\left(\mu^2+3\Delta_s^2\right)\Theta(\mu-\Delta_s)~,
\end{split}
\ee
where $\Delta_\pm=|\Delta_{\rm SO}\pm\Delta_z|$. Using the above expression in Eq \eqref{eq:sigma2}, one can find the charge conductivity at zero temperature, which in the units of $e^2/h$ reads
\be\label{sigma_short}
\sigma~\left[\frac{e^2}{h}\right]=\frac{4\hbar^2 v_{\rm F}^2\mu^2}{n_{imp} V_0^2}\sum_s \frac{\Theta(\mu-\Delta_s)}{\mu^2+3\Delta_s^2} ~.
\ee

\emph{Long-range impurities with TF screening-}
The simplest approximation to include the static screening of charged impurities is the Thomas-Fermi approximation, in which one uses expression~\eqref{eq:v_tf} for the screened electron-impurity interaction.
Using Eq.~\eqref{eq:v_tf} in Eq.~\eqref{eq:taut}, we can obtain an analytic expression for the transport relaxation time
\begin{widetext}
\be\label{eq:tauTF}
\begin{split}
\frac{1}{\tau^s_{\rm tr}(\varepsilon_{\rm F})}
&=\frac{n_{imp} e^4}{32\pi \varepsilon_0^2\hbar k_{{\rm F}s}^2\mu^2}\Theta(\mu-\Delta_s)
\bigg\{ \left\{16k_{{\rm F}s}^4 \pi b_s-40k_{{\rm F}s}^3 q_{\rm TF}+\pi q_{\rm TF}^2(3q_{\rm TF}^2-2\mu^2)(b_s-1)+4q_{\rm TF}k_{{\rm F}s}(3q_{\rm TF}^2-2\mu^2) \right. \\
&\left.+2\Delta\left[4\pi k_{{\rm F}s}^2(2b_s-1)-\pi q_{\rm TF}^2(b_s-1)-4q_{\rm TF}k_{{\rm F}s}\right]
+2\pi k_{{\rm F}s}^2 \left[3q_{\rm TF}^2(2-3b_s)+2\mu^2(2b_s-1)\right] \right\}\\
& -2b_s\left[16k_{{\rm F}s}^4+q_{\rm TF}^2(3q_{\rm TF}^2-2\mu^2)+2k_{{\rm F}s}^2(8\mu^2-9q_{\rm TF}^2)
 +2\Delta^2(8k_{{\rm F}s}^2-q_{\rm TF}^2)\right]\arctan(\frac{2k_{{\rm F}s}}{a_s})\bigg\}~,
\end{split}
\ee
\end{widetext}
where $a_s=\sqrt{q_{\rm TF}^2-4k_{{\rm F}s}^2}$ and $b_s=q_{\rm TF}/a_s$, with $k_{{\rm F}s}$ being the Fermi wave-vector of the $s$ subband. Finally, the analytic form of the conductivity could be obtained from
\begin{equation}
\sigma~[\frac{e^2}{h}]=\frac{\mu}{\hbar}\sum_s \tau^s_{\rm tr}(k_{{\rm F}s}) \Theta(\mu-\Delta_s)~.
\end{equation}


\end{document}